\pdfoutput=1 

\documentclass[a4paper,
               keeplastbox,   
               ]{jacow}
\usepackage{pdfpages,multirow,ragged2e} %
\makeatletter%
	\ifboolexpr{bool{xetex}}
	 {\renewcommand{\Gin@extensions}{.pdf,%
	                    .png,.jpg,.bmp,.pict,.tif,.psd,.mac,.sga,.tga,.gif,%
	                    .eps,.ps,%
	                    }}{}
\makeatother

\ifboolexpr{bool{xetex} or bool{luatex}} 
 {}                                      
 {\usepackage[utf8]{inputenc}}           

\usepackage[USenglish]{babel}

\ifboolexpr{bool{jacowbiblatex}}%
 {%
  \addbibresource{jacow-test.bib}
  \addbibresource{biblatex-examples.bib}
 }{}
\begin{document}

\newcommand{\BVIMIN}{\texttt{B:VIMIN} }
\newcommand{\BVIMINsetting}{\texttt{B\_VIMIN} }
\newcommand{\BVIMAX}{\texttt{B:VIMAX} }
\newcommand{\BIMINER}{\texttt{B:IMINER} }
\newcommand{\BLINFRQ}{\texttt{B:LINFRQ} }
\newcommand{\BVIPHAS}{\texttt{B:VIPHAS} }
\newcommand{\IIB}{\texttt{I:IB} }
\newcommand{\IMDATFORTY}{\texttt{I:MDAT40} }
\newcommand{\IMXIB}{\texttt{I:MXIB} }

\title{Developing Robust Digital Twins and Reinforcement Learning for Accelerator Control Systems at the Fermilab Booster}

\author{D.Kafkes \thanks{dkafkes@fnal.gov}, Fermi National Accelerator Laboratory, Batavia, IL USA 60510 \\ M. Schram, Thomas Jefferson National Accelerator Facility, Newport News, VA USA 23606}


\maketitle

\begin{abstract}

We describe the offline machine learning (ML) development for an effort to precisely regulate the Gradient Magnet Power Supply (GMPS) at the Fermilab Booster accelerator complex via a Field-Programmable Gate Array (FPGA). As part of this effort, we created a digital twin of the Booster-GMPS control system by training a Long Short-Term Memory (LSTM) to capture its full dynamics. We outline the path we took to carefully validate our digital twin before deploying it as a reinforcement learning (RL) environment. Additionally, we demonstrate the use of a Deep Q-Network (DQN) policy model with the capability to regulate the GMPS against realistic time-varying perturbations. 

\end{abstract}

\section{BACKGROUND}

Recently, the challenge and cost of hand-tuning and controlling accelerators has resulted in a push to leverage deep learning ~\cite{7454846,Edelen:2017ewy,yetanotherauraleefelpaper2017,Duris_2020}. In this study, we present continuing work on a real-time artificial intelligence (AI) control system for precisely regulating the Gradient Magnet Power Supply (GMPS), an important subsystem of the Fermilab Booster accelerator complex.

The GMPS is realized as four power supplies, evenly distributed around the Fermilab Booster. Each powers one of four total gradient magnets, which are responsible for steering and accelerating the 400~MeV proton beam the Booster receives from the linear accelerator to 8~GeV ~\cite{rookie,Ryk:1974mu}. The GMPS operates on a 15~Hz cycle between the injection at minimum current and beam extraction at maximum current. Unfortunately, without any regulation, the fitted minimum of the magnetic field may vary from the set point by as much as a few percent, significantly reducing the beam flux available to experiments run at the lab ~\cite{PRABpaper}. This deviation trends with factors such as electrical ground movement, the operation of other nearby high-power radio-frequency systems, and even ambient temperature changes ~\cite{rookie,Ryk:1974mu}.

\begin{figure}
\centering
\includegraphics[width=0.4\textwidth]{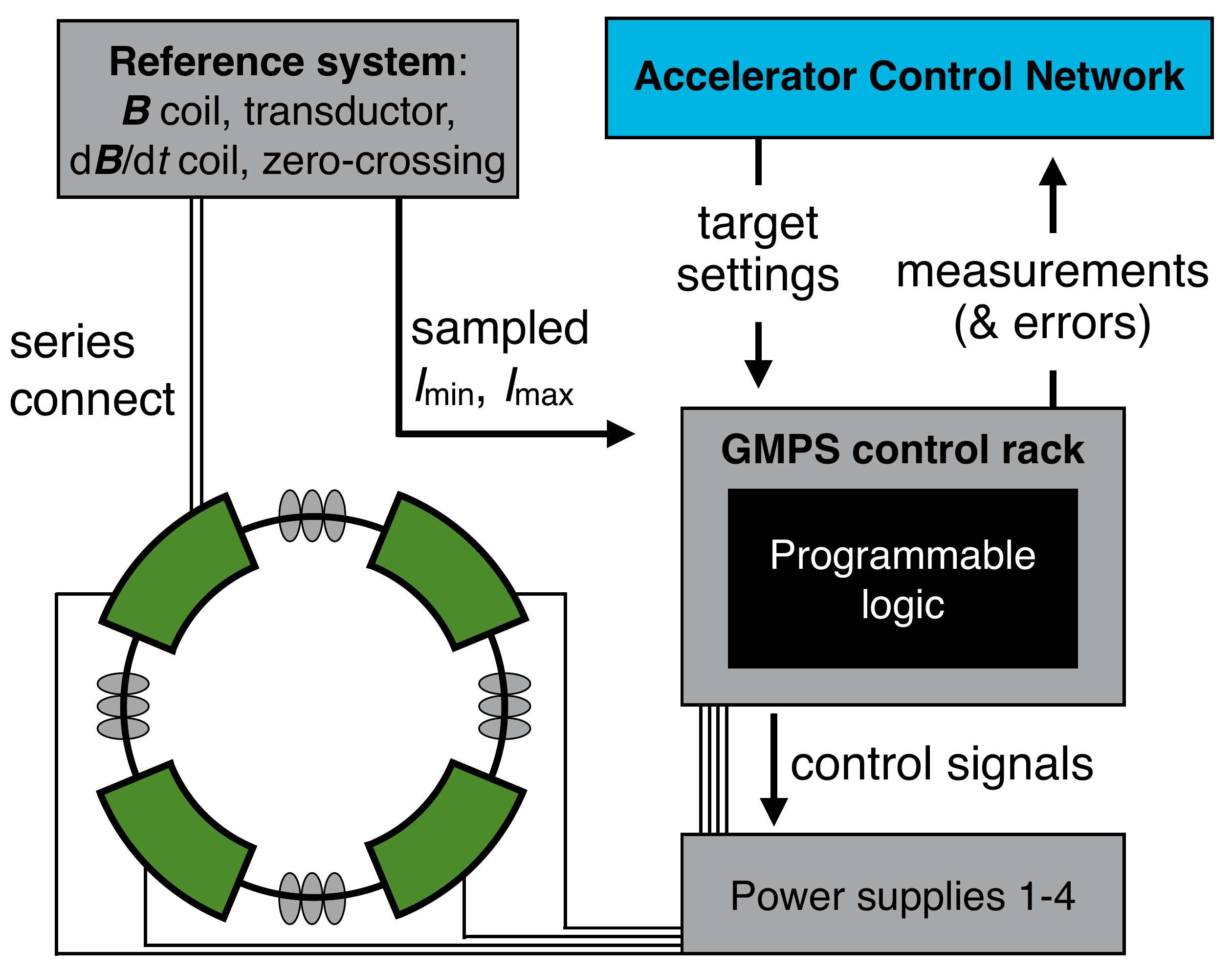}
\caption{Overview of current PID-GMPS control system.}
\label{fig:schematic}
\end{figure}

In order to improve the agreement of the resulting observed minimum and maximum currents with their set points, a proportional-integral-derivative (PID) control scheme applies compensating offsets to the GMPS driving signal as a means of regulation (see Fig. \ref{fig:schematic})~\cite{pid1,pid2}. Presently, a human operator specifies a target program for \BVIMIN and \BVIMAX, the PID-GMPS compensated minimum and maximum currents respectively, via the Fermilab Accelerator Control Network. This signal is then transmitted to the GMPS control board allowing the PID regulator to use the previous 15~Hz cycle to calculate estimates for the minimum and maximum current offset and then uses these values to adjust the power supply program in the current cycle ~\cite{pid1,pid2}.

Presently, this PID-GMPS regulation system achieves errors corresponding to roughly 0.1\% of the set value ~\cite{PRABpaper}. Our ultimate goal is to improve on this error by replacing the PID-GMPS system with a reinforcement learning (RL) approach. This RL-GMPS system will leverage a framework in which an artificial intelligence (AI) agent learns to achieve some end goal through feedback from interactions with its environment ~\cite{sutton_barto_rl2018,Fran_ois_Lavet_2018}. We aim to deploy such an RL agent to control changes in \BVIMIN in order to minimize \BIMINER, the difference between the observed minimum current reading and setting. This agent will ultimately control the GMPS system via a field-programmable gate array (FPGA). However, since training the agent online involves substantial risk, we prototyped our models offline.


\section{METHODS}

We collected approximately six months of time series data from the Booster complex and ultimately selected 250,000 continuous time steps from March 10, 2020 to develop our preliminary algorithms ~\cite{PRABpaper}. Here algorithms refer to both the methods used to build a model that can reliably capture the dynamics of the Booster-GMPS system and to develop a reinforcement learning framework. For complete details on our data collection process and access to our full published dataset please see ~\cite{boostr}.


\subsection{Digital Twin and RL Policy Models}

Our offline ML development involved the training of two different neural network models: a surrogate or digital twin model and the RL agent policy model. Here we implement a simple multi-layer perceptron (MLP) as our agent policy model, the utility of which will be further described in the next subsection, and a stacked Long-Short Term Memory (LSTM) network to capture GMPS dynamics ~\cite{PRABpaper}.

MLPs are standard feedforward neural networks which take in and iteratively feedforward data through many layers of perceptrons (neurons). Each of these neurons involves a function that multiplicatively weights the input vectors, sums them together, adds a bias term; and then applies a non-linear activation function. The chaining of these functions results in a network.


\begin{table}[h!]
\caption{The DQN-MLP policy model architecture.}
\centering
\resizebox{.9\columnwidth}{!}{
\begin{tabular}{r|c|c|c|c}
Layer & Layer Type & Outputs & Activation & Parameters \\
\hline
1 & Dense & 128 & ReLU & 768 \\
2 & Dense & 128 & ReLU & 16,512\\
3 & Dense & 128 & ReLU & 16,512\\
4 & Dense & 7 & Linear & 903 \\
\hline
Total &  &  &  & 34,695 
\end{tabular}}
\end{table}

Beyond MLPs, there are many possible choices of neural network architectures. The one most relevant to capturing the the Booster-GMPS system's multiple frequency modalities in our surrogate model is a type of network known as a Long-Short Term Memory (LSTM)~\cite{PRABpaper}. Unlike standard feedforward networks, during each forward pass, LSTMs are able to learn about previous inputs through the accumulation of weights in a hidden global state variable. This mechanism is useful for modeling time series data, which is exactly what we had collected for this effort.

\begin{table}[h!]
\centering
\caption{We stacked LSTM modules together to create our digital twin architecture.}
\resizebox{.9\columnwidth}{!}{
\begin{tabular}{r|c|c|c|c}
Layer & Layer Type & Outputs & Activation & Parameters \\
\hline
1 & LSTM & 256 & Tanh & 416,768 \\
2 & LSTM & 256 & Tanh & 525,312 \\
3 & LSTM & 256 & Tanh & 525,312 \\
4 & Dense & 3 & Linear & 771 \\
\hline
Total &  &  &  & 1,468,163 
\end{tabular}}
\end{table}




\subsection{DQN Reinforcement Learning}

Reinforcement learning is a training framework that involves an AI agent interacting with an environment to maximize a defined reward over many fixed-iteration-length episodes~\cite{sutton_barto_rl2018,Fran_ois_Lavet_2018}. The agent's actions within this environment are defined by a policy model. As stated above, we used an MLP for this policy model in accordance with the deep $Q$-network (DQN) approach, which trains this neural network to learn the action-value function--- $Q$-value--- that maps a discrete number of agent actions to rewards~\cite{dqn,mnih2013playing}. To keep our action space finite, we discretized the \emph{change} of \BVIMIN using steps of just seven different sizes, including an option for zero-size change ~\cite{PRABpaper}.


At each time step $t$, our surrogate model environment takes in the control action $A_t$ determined by the RL agent MLP-DQN policy model as the small compensation to be applied to \BVIMIN based on the current state $S_t$. The digital twin then provides the new system state $S_{t+1}$ along with an associated reward $R_{t+1}$. In our studies, the state is composed of the variables inputted to the surrogate model (discussed at length in the section below) and the reward is calculated from the \BIMINER output by the surrogate:
\begin{equation}
R_t = -|\BIMINER(t)|\,.
\label{eqn:reward}
\end{equation}
\noindent Optimizing the agent's policy actions over the training horizon is defined to mean maximizing the long-term integrated reward, which is calculated over each fixed-length episode. 


\section{DIGITAL TWIN DEVELOPMENT AND VERIFICATION}

In our preliminary result ~\cite{PRABpaper}, we formulated our stacked LSTM surrogate model to capture the dynamics of:
\begin{multline}
\BVIMIN + \BIMINER + \BLINFRQ + \IIB + \IMDATFORTY \rightarrow \\
\BVIMIN + \BIMINER + \BLINFRQ.
\end{multline}
\noindent Here \BLINFRQ is the measured offset from the 60~Hz line frequency, and \IIB and \IMDATFORTY provide measurements of the main injector bending dipole current through different communication channels. This model was trained using MinMax scaling~\cite{scikit-learn} and a 150 step lookback, i.e. 150 previous timesteps of the input variables were fed into the model in order to predict the next timestep forward in the output variables~\cite{PRABpaper}. Here we describe the validation process we used to verify and improve upon our initial result.

First, we distilled our surrogate model into the simplest possible combination of variables we aimed to regulate: \BVIMIN + \BIMINER $\rightarrow$ \BVIMIN + \BIMINER. From this most basic model, we explored using a much smaller lookback window of 15 timesteps as well as the use of Robust scaling~\cite{scikit-learn}. Since we found the 15 timestep lookback and MinMax scaling to be performant, we ultimately decided to keep the original scaling and move forward to experiment with this much shorter lookback window.

After iterating over this most basic model, we began the forward selection process, experimenting with the inclusion of different variables. The variables we considered for these studies included the inputs from the original model: \BVIMIN, \BIMINER, \BLINFRQ, \IIB, and \IMDATFORTY; as well as \BVIMINsetting (the GMPS minimum current set point), \BVIMAX (the compensated maximum GMPS current), \BVIPHAS (the GMPS ramp phase with respect to line voltage), and \IMXIB (the main injector dipole bend current). \BVIMINsetting was included on the suggestion of subject-matter experts ~\cite{rachaelreport}; and the three additional variables were selected based on the results of a Granger Causality study ~\cite{granger}. For details on how this analysis was performed, please see ~\cite{PRABpaper}.


The results of these surrogate model experiments are presented in the table below, which displays the final loss attained by the model with the given configuration. After comparing results, we decided to move forward with the 6 to 2 model since including the other three variables in the 9 to 2 model made only a slight difference in the final loss:
\begin{multline}
\BVIMIN + \BIMINER + \BVIMINsetting + \BLINFRQ \\ + \IIB + \IMDATFORTY \rightarrow \BVIMIN + \BIMINER
\end{multline}
\begin{table}[h!]
\caption{Training Mean-Squared Error (MSE) results from digital twin experimentation studies.}
\centering
\resizebox{.9\columnwidth}{!}{
\begin{tabular}{c|c}
\hline
Model              & MSE ($10^{-6}$)  \\ \hline
\BVIMIN + \BIMINER $\rightarrow$ \\ \BVIMIN + \BIMINER & 449.0567 \\ \hline
\BVIMIN + \BIMINER + \BVIMINsetting $\rightarrow$ \\ \BVIMIN + \BIMINER & 379.5542 \\ \hline
\BVIMIN + \BIMINER + \BLINFRQ \\ + \IIB + \IMDATFORTY $\rightarrow$ \\ \BVIMIN + \BIMINER & 346.6192 \\ \hline
\BVIMIN + \BIMINER + \BVIMINsetting \\ + \BLINFRQ + \IIB + \IMDATFORTY \\ $\rightarrow$ \BVIMIN + \BIMINER & 314.3544 \\ \hline
\BVIMIN + \BIMINER + \BVIMINsetting \\ + \BLINFRQ + \BVIMAX + \BVIPHAS \\ + \IIB + \IMDATFORTY + \IMXIB \\ $\rightarrow$ \BVIMIN + \BIMINER & 294.6336 \\ \hline
\end{tabular}}
\end{table}

Additionally, we tried decomposing the variables from the models mentioned above into signal and noise vectors using Empirical Mode Decomposition~\cite{EMD}. Despite the fact that this resulted in marginally better performance, we decided that this would be too difficult to implement in real-time on the board funneling input data to the FPGA. For this reason, we omit these results here. Similarly, after completing the digital twin validation and verification process, we decided to create our own version of the MinMax scaler rather than using the transformation available to us via the scikit-learn library since using this premade scaler could not be easily implemented on the FPGA-side.

\subsection{Uncertainty Quantification}

In order to provide a prediction with statistical interoperability, we performed concrete dropout as a means of uncertainty quantification ~\cite{CD}. The concrete dropout process involves introducing tunable uncertainty into a network training process through the addition of a dropout layer, which randomly removes inputs to the following layer with some probability $p$ at each forward pass. This causes the training of the network's other weights and biases to adjust without these ``dropped out" neurons. Once this dropout layer has been added to the network, $p$ can be adjusted to take on a different value during inference.

After training our surrogate with a dropout layer inserted after the first LSTM, we set the layer to probabilities ranging from $[0.05, 0.1, 0.15, 0.2, 0.25, 0.3, 0.5]$, and used the inferred outputs of our network to match the actual underlying distribution of the data. We found a value of $p = .2$ gave us our best results: reconstructing the modeled distribution of \BVIMIN at $103.3930 \pm 0.0297$ (underlying distribution: $103.3940 \pm .0314$); and \BIMINER at $0.0012 \pm 0.2090$ (underlying distribution of $.0011 \pm .2181$).


\section{Preliminary RL Results}

\begin{figure}[h!]
\centering
\includegraphics[width=0.5\textwidth]{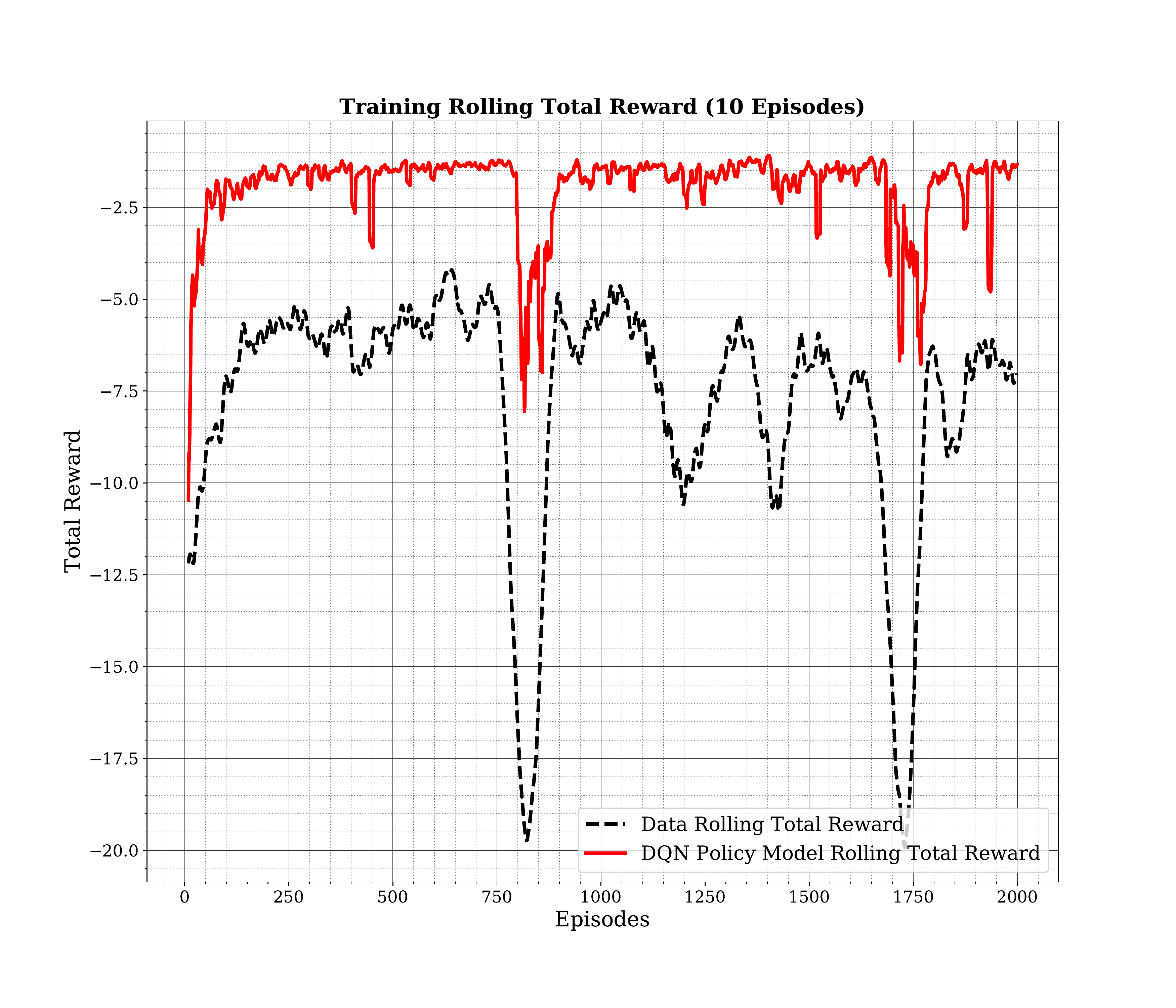}
\includegraphics[width=0.5\textwidth]{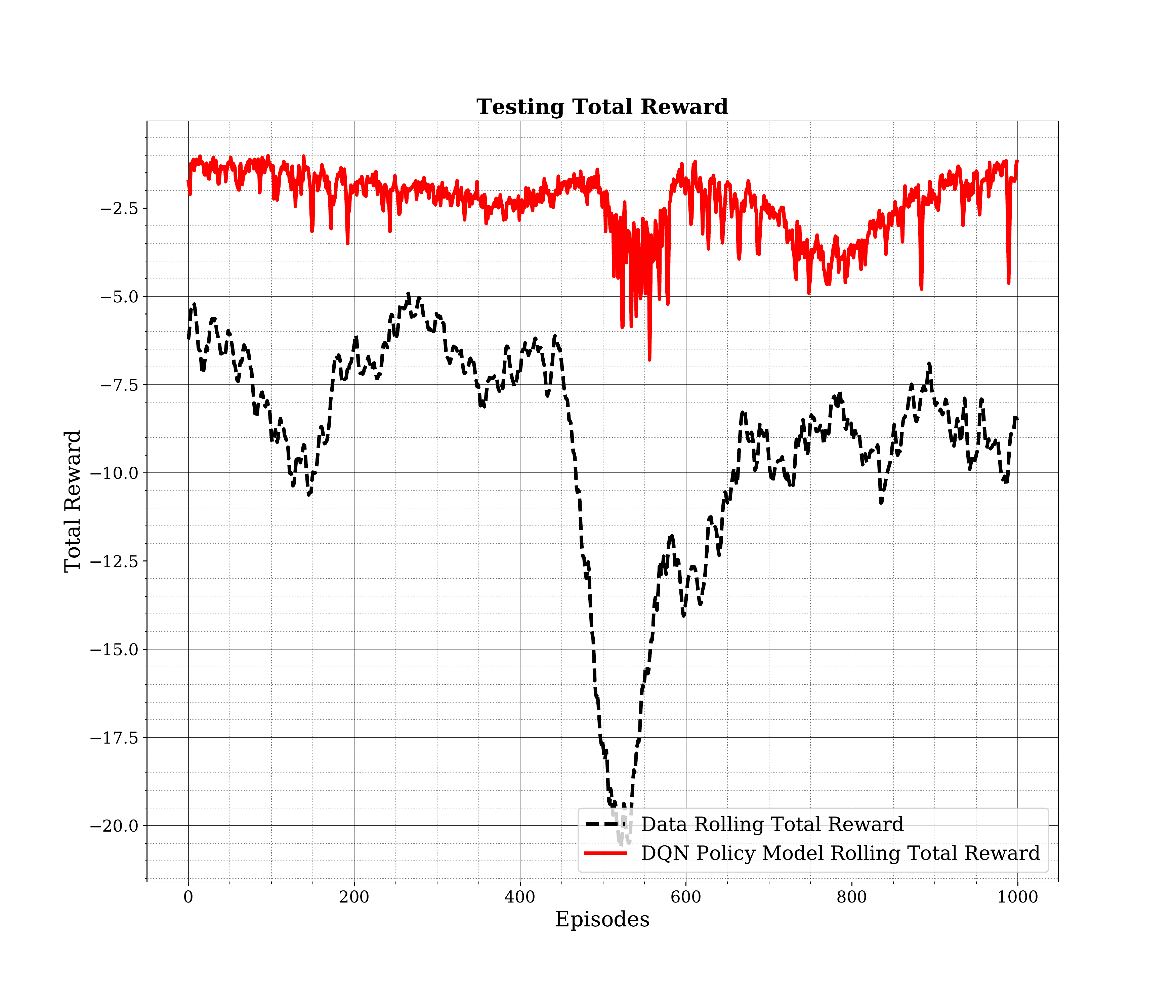}
\caption{Results of training (top) and testing (bottom).}
\label{fig:trainingandtesting}
\end{figure}

Finally, we present our most recent RL results, training and deploying our trained MLP-DQN policy model within our verified digital twin environment in Fig. \ref{fig:trainingandtesting}. When comparing the DQN-GMPS system results to the PID-GMPS controller, we see a factor of 2-4x improvement.



\section{CONCLUSION}

We outlined the steps we took to carefully validate our digital twin of the Booster-GMPS system--- perhaps the most important aspect of our offline machine learning development. After all, without a robust surrogate model to support training, we would not be able to trust the deployment of the trained agent on the live system via an FPGA in the future.

\section{ACKNOWLEDGEMENTS}

This proceedings was created as part of the ``Accelerator Control with Artificial Intelligence'' Project conducted under the Fermilab Laboratory Directed Research and Development Program (Project ID \textit{FNAL-LDRD-2019-027}). The manuscript has been authored by Fermi Research Alliance, LLC under Contract No. DE-AC02-07CH11359 with the U.S. Department of Energy, Office of Science, Office of High Energy Physics and is registered at Fermilab as Technical Report Number \textit{FERMILAB-CONF-21-230-AD-SCD}.
%
%
	


\end{document}